\documentclass[fleqn,10pt]{wlscirep}
\usepackage[utf8]{inputenc}
\usepackage[T1]{fontenc}

\usepackage{graphicx}
\usepackage{caption}
\usepackage{subcaption}
\usepackage{amsmath}
\usepackage{todonotes}
\usepackage{comment}
\usepackage{float}

\newcommand*{\figref}[2][]{%
  \hyperref[{fig:#2}]{%
    ~\ref*{fig:#2}%
    \ifx\\#1\\%
    \else
      #1%
    \fi
  }%
}

\author[1,+]{Jolien Cremers}
\author[1,2,+]{Benjamin Kohler}
\author[1,3,+]{Benjamin Frank Maier}
\author[1]{Stine Nymann Eriksen}
\author[1,2]{Johanna Einsiedler}
\author[1]{Frederik Kølby Christensen}
\author[2,3]{Sune Lehmann}
\author[2,4]{David Dreyer Lassen}
\author[1,5,6]{Laust Hvas Mortensen}
\author[2,4,*]{Andreas Bjerre-Nielsen}
\affil[1]{Methods and Analysis, Statistics Denmark, Copenhagen, 2100, Denmark}
\affil[2]{Center for Social Data Science (SODAS), University of Copenhagen, Copenhagen, 1353, Denmark}
\affil[3]{Department of Applied Mathematics and Computer Science, Technical University of Denmark, Kongens Lyngby, 2800, Denmark }
\affil[4]{Department of Economics, University of Copenhagen, Copenhagen, 1353, Denmark}
\affil[5]{Department of Public Health, University of Copenhagen, Copenhagen, 1353, Denmark}
\affil[6]{The Rockwool Foundation, Copenhagen, 1472, Denmark}

\affil[*]{abn@sodas.ku.dk}

\affil[+]{These authors contributed equally to this work}

\title{Unveiling the Social Fabric: A Temporal, Nation-Scale Social Network and its Characteristics}

\begin{abstract}{
  Social networks shape individuals' lives, influencing everything from career paths to health. This paper presents a registry-based, multi-layer and temporal network of the entire Danish population in the years 2008-2021 ($\sim$7.2 mill.). Our network maps the relationships formed through family, households, neighborhoods, colleagues and classmates. We outline key properties of this multiplex network, introducing both an individual-focused perspective as well as a bipartite representation. We show how to aggregate and combine the layers, and how to efficiently compute network measures such as shortest paths in large administrative networks.
  Our analysis reveals how past connections reappear later in other layers, that the number of relationships aggregated over time reflects the position in the income distribution, and that we can recover canonical shortest path length distributions when appropriately weighting connections.
Along with the network data, we release a Python package that uses the bipartite network representation for efficient analysis.}
\end{abstract}

\begin{document}
\flushbottom
\maketitle
\thispagestyle{empty}

\section{Introduction}

Interactions between individuals are at the core of social life. Consequently, researchers across many different fields have long been interested in understanding the structure of these interactions, i.e., social networks \cite{watts_new_2004,barabasi_network_2005, jackson_social_2011,wasserman_social_1994}.
This is an inherently challenging task as social interactions are rarely directly observed by researchers and are difficult to reproduce in an experimental setting. Thus, until fairly recently our understanding of social network structures was based on surveys where participants are asked to self-report with whom they interact \cite{eagle_inferring_2009,marsden_network_1990,bernard_informant_1982}. This type of data collection is, however, limited in scope and prone to multiple types of bias.

Within the last two decades, the advent of digital media and mobile phones has led to the possibility of deriving graphs from digital trace interaction data, allowing researchers to examine these structures at scale and without inherent self-reporting bias\cite{stopczynski_measuring_2014,ugander_anatomy_2011,cha_measuring_2010,leskovec_learning_2012,eagle_inferring_2009,onnela_structure_2007}. However, these social networks derived from media platforms often suffer from algorithmic bias, platform-specific behavior, and self-selected, non-random populations. These factors limit the generalizability of insights into an improved understanding of broader social structures \cite{bokanyi_anatomy_2023}.

At the same time, continuous improvements in the granularity and scope of administrative data collection have made it possible to establish network structures based on real-life social foci, i.e., entities around which joint activities are organized \cite{feld_focused_1981}. Recently, van der Laan et al. \cite{van_der_laan_whole_2023} created a network of the entire population in the Netherlands by linking individuals through common households, neighborhoods, companies, educational institutions, and family relationships. While these types of networks do not capture realized social interactions, they depict the current opportunity structures for forming relationships on a national scale. This means that two individuals that are connected via any (or multiple) of those layers, have a much higher likelihood of being connected to each other in real life as opposed to any random pair in the data. The use of registry data presents immense possibilities as next to its massive scale, unselected nature and the inclusion of several social foci, registry data allows the addition of a wide variety of individual-level socio-economic and health-related information to the network. Consequently, these network data can aid the study of large-scale social phenomena and epidemiological questions. 
 Importantly, individuals change their social foci throughout their lifetime, e.g., they change workplace or address. As a result, network structures are highly dynamic and path-dependent. Considering that connections formed in the past, such as former classmates and colleagues, continue to hold relevant social information, it is important to incorporate a temporal dimension when studying social networks. One limitation of existing approaches leveraging administrative data for this purpose is their focus on a single point in time\cite{van_der_laan_whole_2023}. 
 %
%
%
A further challenge is the increasing amount of computational resources needed to analyze networks based on vast amounts of data, especially when relying on representations where edges indicate connections between individuals only \cite{ugander_anatomy_2011,van_der_laan_whole_2023}. Here, we see a massive opportunity for improvement by leveraging the bipartite and multilayered nature of administrative network data for measuring social networks \cite{lehmann_biclique_2008}.
%

In this paper, we present a nation-scale network for the entire population of Denmark derived from registry data for every year between 2008 and 2021. To address the \textit{first} challenge of the missing temporal dimension, we introduce the concept of time-span, i.e., including edges from multiple years of administrative social networks. We use this framework to investigate how the network's structure changes over different time horizons, conceptually akin to assuming different durability scopes of individual relationships.
We address the \textit{second} challenge of computational intractability by developing a novel representation of administrative networks that reflects the inherent structure of the network. Relationships in administrative networks are formed over shared foci, such as common workplaces, school classes, or nearby addresses \cite{feld_focused_1981}. 
In this picture, edges can also be viewed as individual's relations to these foci, e.g., working for a specific employer in the workplace layer. We discuss how this bipartite view on administrative networks leads to a massive reduction of edges and, thus, to an increase in performance for many commonly used network algorithms. A further challenge in using administrative networks is that it is not clear to what extent people communicate in different layers, e.g., with people in their neighborhoods compare to say colleagues or classmates. Our computational framework allows for weighting edges to more plausibly reflect distances between individuals in these networks.

In our analysis, we show that the temporal stability of edges varies substantially depending on the nature of the relation. As a consequence, the level to which nodes accrue edges over time also differs, leading to increasing layer-level degree inequality. The addition of the temporal dimension also leads to reduced average shortest-paths when considering individual layers.  Using the weighting scheme described in Subection~\ref{sec:shortest_paths} to adjust for the unrealistic inflation of edges, combining the different layers into one large network yields a shortest path distribution similar to other realized social networks.
Furthermore, we show that the extent to which the degree reflects income levels for different age cohorts increases with the introduction of time-span. This finding illustrates how the properties of the network and the added time dimension encode actual life outcomes of individuals.

The network, as well as the software to create and utilize the novel representation, is available to all researchers with access to research accounts at Statistics Denmark (see how to access the data in Subsection~\ref{sec:access} and the use of the software in Subsection ~\ref{sec:software}) \cite{maier_regnet_2024-1}.

\section{Construction of a temporal, nation-scale network}
This section delineates the construction of a temporal, nation-scale network based on registry data, outlining the multi-dimensional structuring of nodes and edges to create diverse subnetworks for dynamic and layered analysis.
Network science defines networks in terms of graphs, consisting of entities referred to as \textit{nodes} and links between them, referred to as \textit{edges}. 
In social networks, nodes frequently represent individuals and the edges describe some type of social interaction, e.g., a friendship on a social media platform. 
The network we construct is not based on a single definition for all nodes and edges but differentiates along multiple dimensions -- two different views, time and the relation type also referred to as layer. We will introduce each of these dimensions in the following.

\subsection{A dual view on the same information: Individual-centered and bipartite}

We introduce several fundamental, commonly shared foci as layers in a nation-scale network based on registry-data, in particular families, educational settings, workplaces, neighborhoods, and households \cite{feld_focused_1981}. We adopt a dual view on the individuals and their shared foci settings. In the first (unipartite or individual-centered view), we focus on individuals and abstract away the foci via which they can form interactions. In this perspective, nodes represent individuals and edges their relation within the respective layer. For example, an edge in the neighborhood-layer represents a neighbor-relationship between the two connected nodes. 

The second perspective (bipartite view) acknowledges the inherent bipartite structure of the network layers, i.e., relations in the network are formed via concrete foci settings. For example, work relationships emerge through common workplaces, households and neighborhoods through sharing the same or a close-by address. In the bipartite view we explicitly include these foci settings as additional nodes (henceforth also referred to as container nodes). For example, each workplace becomes such a container node. In this bipartite perspective no edges between individuals exist -- only between individuals and container nodes. As individuals are nonetheless implicitly connected through shared container nodes, this perspective yields the same relationship information but comes with several computational and conceptual advantages, which we discuss in Subection \ref{sec:structure_properties}. Information for the individual-centered view are provided via edge-lists, while information for the bipartite-view is provided via a bipartite network file containing all individuals in the population and their household, address, workplace and institution (school) information.

\subsection{Temporal dimension}

The network encompasses data on registered residents of Denmark spanning from January 1, 2008, to December 31, 2021. Edges across the five distinct layers are updated and documented with annual granularity, reflecting changes and continuities in social connections over time.

\subsection{Layer dimension}
The definition of when an edge is said to exist between a pair of nodes differs per layer and we outline these definitions below. More detailed information about all layers can be found in the Appendix \ref{sec:appendix-layers}.

\subsubsection{Family layer}
Edges in this layer consist of first and second degree family relations. Family relations are constructed on the basis of child-parent relationships. We include both biological and adoptive relationships. Note that the approach we take here follows the one in the Dutch administrative network\cite{van_der_laan_whole_2023}. The relations present in this layer consist of child, grandchild, half sibling, full sibling, sibling (unknown), cousin, co-parent, parent, grandparent, aunt/uncle and niece/nephew. Where sibling (unknown) indicates that the children share at least one parent, but information on one or both of the other parents is missing. This layer only exists in the individual-centered view as the nature of the relation depends on the position in the network and is thus not inherently bipartite.

\subsubsection{Household layer} \label{description-hh}
Edges in this layer indicate membership of the same household on the first of January of a specific year. A household consists of one individual or a couple with or without children that live at the same address. A more detailed description of the household definition used can be found in Appendix ~\ref{sec:appendix-layers}. In the bipartite view, the household is included as a container node.

\subsubsection{Neighborhood layer}
In the neighborhood layer, we include edges between an individual's and the members of the geographically closest ten households on the first of January of a specific year (as defined in the household layer) within a distance of 50 meters. This is the same approach as in the construction of the Dutch administrative network \cite{van_der_laan_whole_2023} take. When multiple households exist within the same distance, households are randomly selected. In the bipartite view, addresses are added as container nodes and household container nodes are connected to these addresses. Neighbors are defined as edges between these address containers. Therefore this layer is not strictly bipartite as addresses are connected to each other.

\subsubsection{Colleague layer}
In this layer, we use the information on work location from the labor market registry \cite{petersson_danish_2011}. This registry contains a yearly status indicating the place(s) of work on the 30th of November of all individuals present in the Danish population on the 1st of January of the same year. For small workplaces ($\leq$ 100 employees) we include edges between all employees in the yearly edge-list. For large workplaces (>100 employees) we take a random sample of size 100 to create edges between colleagues for the individual-centered view. For the bipartite-view we do not implement such a sampling as here the number of edges grows linearly in the number of colleagues compared to the quadratic growth in the individual-centered view. 

\subsubsection{Classmate layer}
In this layer we use information on enrolled students in primary, secondary, and tertiary education from the Population Education Register \cite{jensen_danish_2011} to create edges between students enrolled at the same school for the same study program in the same grade or year. The data contain information from the 1st grade of primary school (from the year an individual turns six) onward. For the bipartite-view, study programs are included as container nodes to which individuals are connected.

\subsection{Additional data}
The network allows the integration of further individual-level attributes from the Danish national registries to nodes. 
The Danish national registries compose one of the world's largest data repositories. They include, for example, information on labor market affiliation, personal income, transfer payments, education, hospital treatment, healthcare costs as well as redeemed prescription medicine. Available information from the various registries can be linked to individuals using a personal identification number that originates in 1968 \cite{Pedersen2011-an}. From that time onward an increasing number of registries have been established, among others the education registry in 1977 \cite{jensen_danish_2011}, the registry on personal labor market affiliation in 1980 \cite{Petersson2011-xc} and the Danish National Patient Register in 1977 \cite{Lynge2011-et}.  This means that individual-level information on several socio-economic factors as well as health variables can be obtained from more than 50 years back in time.
Furthermore, the bipartite view enables the addition of attributes to the container nodes, including information about the educational institution, employing firm, or address, 
information that in the individual-centric view would have to be tied to edges, thereby drastically increasing the necessary amount of computational memory.

\subsection{Data access}\label{sec:access}

The network is available to researchers through Statistics Denmark by means of several bipartite tables and edge lists. For each layer one bipartite table is available. One exception is the family layer which is not a bipartite structure, but only exists as a table including all family relations (past and present) known up to and including 2022 (see also the description of the family layer in Appendix \ref{sec:appendix-layers}). Note that relations to deceased individuals are included in this table as well (e.g. grandparents). In addition, edge lists are provided for each layer. For the work layer two edge lists are provided one for the smaller workplaces and one for the larger workplaces in which pairs of colleagues have been sampled uniformly at random. Edge lists are created on a year-by-year basis in which we only include edges if both individuals were present in the Danish population for at least one day in a specific year.  For a general description of the rule of access, we refer the reader to the homepage of Research Services at Statistics Denmark \cite{statistic_denmark_data_nodate}. Contact the corresponding author for more information about data access and information about the specific tables and variables.

\section{The network's structure and its properties} \label{sec:structure_properties}

As the nation-scale network is multiplex, spanning multiple years and exists in two different structures, understanding its fundamental properties is a multidimensional endeavor. 
Embracing this multi-dimensionality, we focus on the following aspects: i) how properties evolve over time for single layers, ii) emerging properties when combining layers and iii) the role of the bipartite vs.\ individual-focused representation of the network.

To understand the influence of time on the network's properties, we introduce the concept of time-span. The time-span describes the number of included years prior to the most recent year (2021). For example, a time-span equal to 5 means that all relations between 2016 and 2021 (inclusive bounds) are included in the network. From a social science perspective, time-span could also be interpreted as the individuals' ability to remember past connections.

Before diving into the description of the network's anatomy, it is important to clarify how edges should be interpreted. The network describes relationships in terms of affiliations to some shared opportunity spaces. These spaces are very different in nature and so is their inherent likelihood of representing an actual interaction between two individuals via the shared opportunity space. While for example interactions within households can be seen as guaranteed and regular interactions via family members are plausible, this is to a much lesser degree the case when considering colleagues in a large workplace with thousands of employees. Shared relations can therefore be viewed as what has been described as shared social foci \cite{feld_focused_1981}. These foci thus pose a social opportunity space \cite{bokanyi_anatomy_2023}, which make a realized interaction to a varying degree more likely than interactions to any random individual. We will account for these varying likelihoods when weighting edges to compute shortest paths.

As the network spans an entire population of roughly 7.2 million individuals over multiple years and layers and individuals are by the definition of the layers highly clustered, it contains a substantial number of edges (Figure \figref[A]{grid_descr}). The number of edges varies substantially between layers. For example, households connect only a small set of individuals (4.5 million for a single year), compared to relations formed among colleagues which sum up to 598 Million for a ten year time-span (note the log-scale in Figure~\figref[A]{grid_descr}). The cause for the increased efficiency of the bipartite view for various computations becomes apparent when considering the difference of the number of edges between the two views. By connecting individuals via shared containers, the bipartite view drastically decreases the number of edges, especially for the large colleague and classmate layer for which the bipartite view only contains a fraction of the number of edges of the individual-centered view. This reduction goes up to 98\% when comparing the classmate layer with a ten year time-span of the bipartite to the individual-centered view. The only exception here is the household layer because in the bipartite view single households also have a single edge to an individual container. 
The diminished number of edges in the bipartite view is accompanied by a relatively small increase in the total number of nodes caused by the addition of the container nodes (Figure \figref[C]{grid_descr} - note the linear scale). This results in relevant efficiency gains for most graph algorithms, for example shortest paths. Consider the Dijkstra algorithm with time complexity is at best $\mathcal{O}((V\log V) + E$ \cite{fredman_fibonacci_1984} where the number of nodes is denoted $V$ and the number of edges $E$. As in the bipartite view the number of edges decreases strongly disproportionate to the increase in the number of vertices for all time-spans and layers except households, using the bipartite view yields a substantial efficiency gain. 
Moreover, the bipartite view captures more information than the individual-centered, unipartite one. To illustrate this, consider an example in the colleague layer and with multiple years time-span. Here, we can not infer from three neighboring nodes ($a$, $b$, and $c$) whether they are connected via the same or three separated workplaces where the first workplace connects nodes $a$ and $b$, the second nodes $b$ and $c$, and the third nodes $a$ and $c$. Therefore, the bipartite view cannot be reconstructed from the unipartite projection \cite{lehmann_biclique_2008}. 
Finally, apart from the computational and conceptual advantages described above, the bipartite view also allows researchers to create their own edge definitions.

Next to the number of depicted relations, the layers also vary in their coverage of the total population, i.e., the number of unique nodes (Figure \figref[C]{grid_descr}). Neighborhood, family, and household information exists for a large share of the overall population, as they are independent of age in contrast to colleagues and classmate relations. The introduction of time-span leads to an increase of the number of nodes because individuals who graduated (classmate layer), moved to another country (neighborhood and household layer), retired (colleagues layer), or have deceased in the past are included, as well.

\begin{figure}
    \centering
\includegraphics[width=1.0\textwidth]{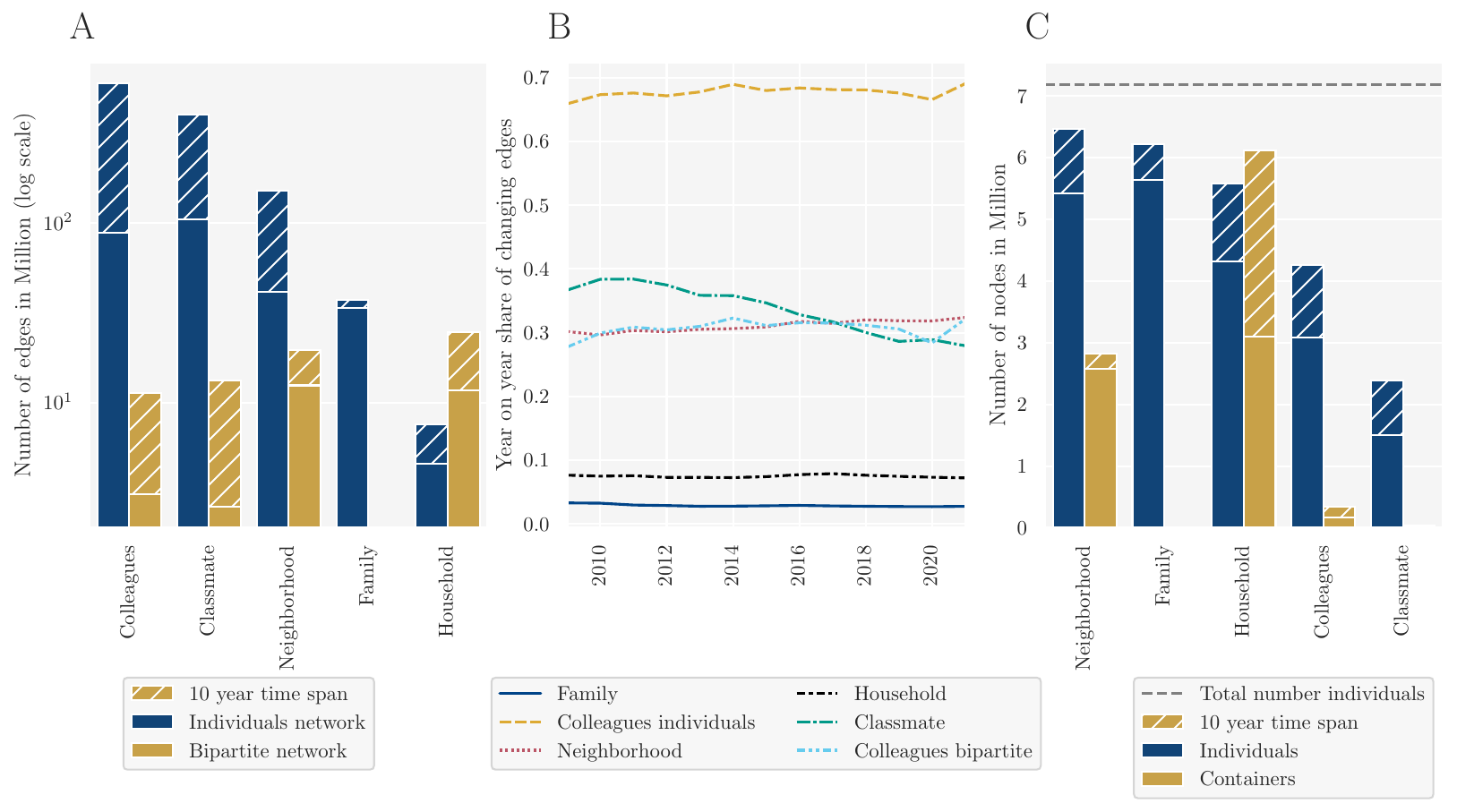}
\caption{
\textbf{Size and Temporal Stability Of The Network.}
Panel A shows the total number of edges per layer for a single year (i.e., 2021) and a ten year time-span for the bipartite and the individual-centered representation of the network (log scale). Panel B depicts the year-over-year share of changing edges, i.e., the number of edges in one year that also exist in the next year relative to the total number of edges. Panel C shows the total number of nodes per layer for a single year and ten year time-span for the individual-centered and the bipartite representation of the network. Note that the family layer does not include bipartite nodes and that there are too little classmate container layers to be visible.}
    \label{fig:grid_descr}
\end{figure}

\subsection{Overlaps and evolution of network layers}

The increase in the number of edges in time-span depends not only on the number of nodes but also reveals information of the stability of the underlying relation. Figure \figref[B]{grid_descr} depicts the year-over-year share of changing edges for different layers. While a single digit percentage of family and household relations change every year, neighborhood, job and participation in educational programs fluctuate more frequently over time. There is also a substantial difference for the colleagues between the views because of the noise added by the sampling in the individual-centered view. \\

\begin{figure}
    \centering
\includegraphics[width=1\textwidth]{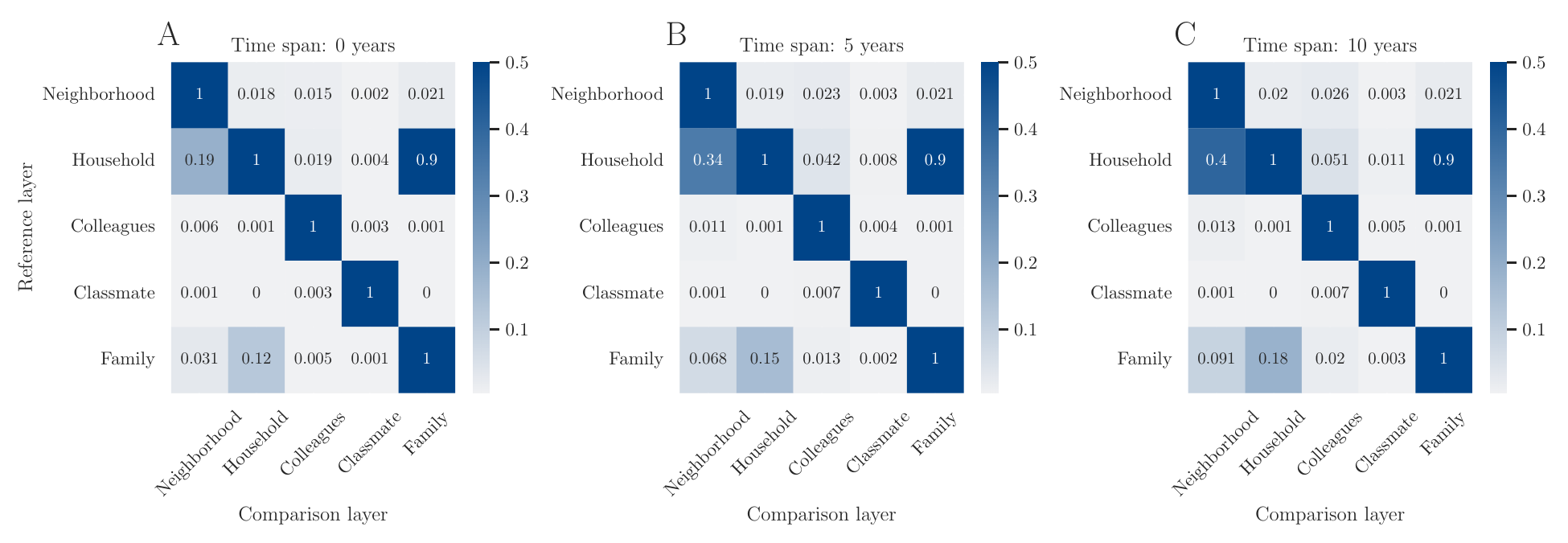}
\caption{\textbf{Edge Overlap With Varying  Time-Span.}
Share of overlapping edges between reference layer (y-axis) and comparison layer (x-axis). The reference layer is measured for the year 2021 for all the plots. The comparison layer includes a time-span of 0, 5, or 10 years before 2021. The overlapping share is computed as the number of overlapping edges divided by the total number of edges of the reference layer. In Panel A, the comparison layer includes no time-span, in Panel B a five-year time-span, and in Panel C  a ten year time-span.}
    \label{fig:overlap_edges}
\end{figure}

In our multiplex network, the share of overlapping edges between layers reveals how much additional information each layer provides \cite{de_domenico_more_2023}. Moreover, it can be argued that overlapping edges can indicate tie-strength as individuals sharing multiple foci points are more likely to also have realized interactions. Figure \ref{fig:overlap_edges} shows the share of overlapping edges between layers, with the y-axis denoting the layers for a single year and the x-axis depicting the layers with the indicated time-span. The heatmap is not symmetric as the baseline for the share is the layer on the y-axis. 
We observe relatively large overlaps between the household, neighborhood, and family layer, which are partly explained by the definition of the layers, as the household layer is defined as a family-like structure, and neighbors with the same address can become households when they meet one of the household criteria (e.g., become parents). More relevant are the relatively small but in time-span increasing overlaps between all other layers. For example, we can see how workplaces have been a relevant matching foci for future households: 5.1\% of all household edges have also been an edge in the colleague layer over the previous ten years. Overall, the small share of overlapping edges underlines how each layer captures a unique aspect of the social sphere. 

\subsection{Node degree distribution}
\begin{figure}
\centering
\includegraphics[width=1.1\textwidth]{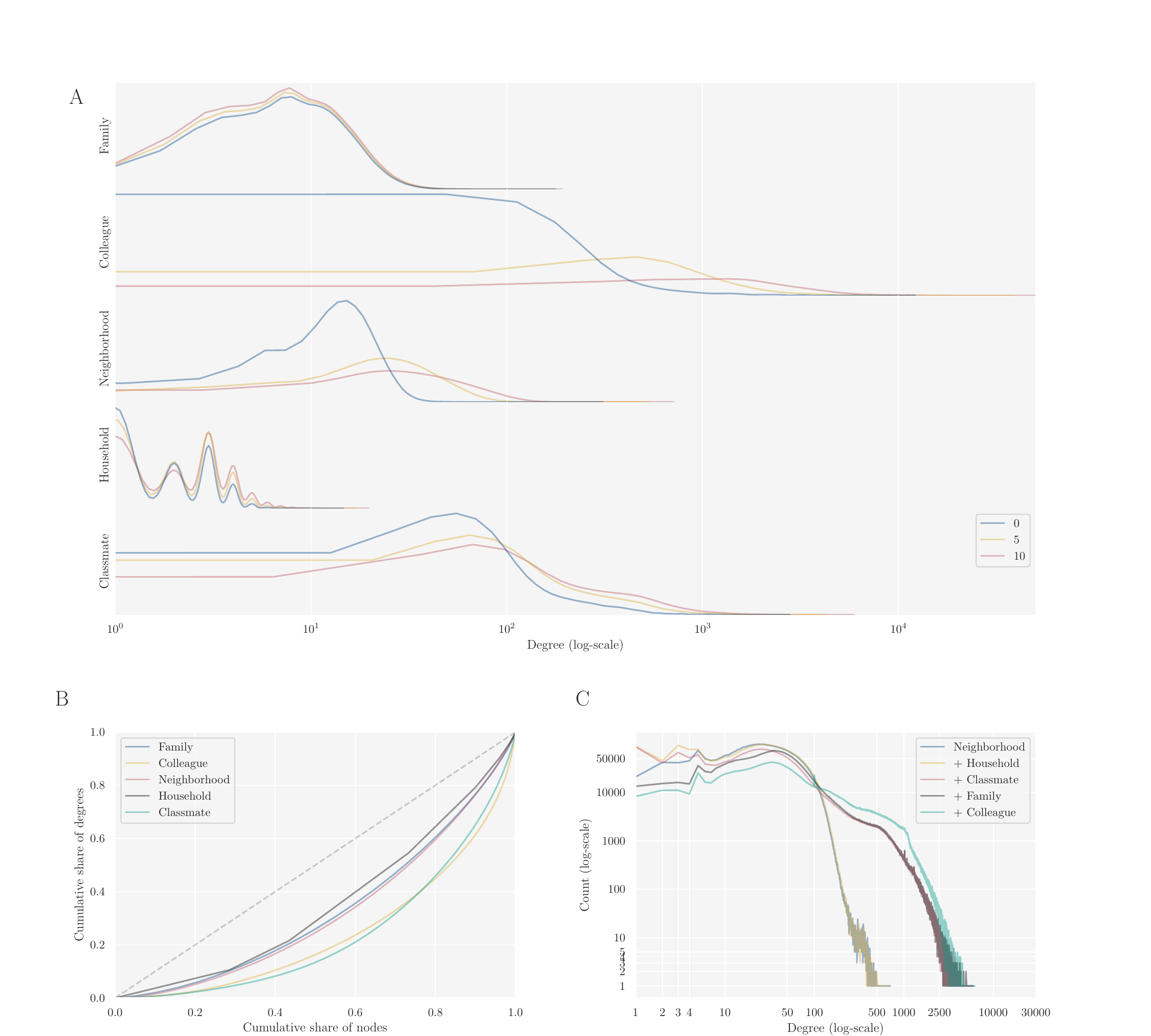}
    \caption{ \textbf{Degree Distribution.}
    Panel A shows the kernel density estimation for the degree distribution for single year (i.e., 2021), five and ten year time-span (log scale). Panel B depicts the Lorenz-curve (cumulative share of degrees for lower x-\% of nodes) by layer. Panel C shows the degree distribution (log-scale) for stacked layers}
    \label{fig:grid_degree_wo_distribution}
\end{figure}

The number of connections of a node is an important indicator of a node's centrality and relevance.
Figure \ref{fig:grid_degree_wo_distribution} depicts three different views on the degree distribution. Panel \figref[A]{grid_degree_wo_distribution} shows the kernel density estimation of the degree distribution depending on layer and time-span. The number of connections naturally vary between layer-types, also becoming apparent by the use of the log-scale on the x-axis. While, e.g., in the household layer 90\% of  individuals are connected to five or less other individuals for a ten year time-span, this number is 5915 for connections via the colleague layer. As increasing the time-span adds additional relations, the degree distributions shift upwards for all layers with respectable variation between the family or household and the neighborhood, colleagues or classmate layer. As observed in many other real-world social and artificial networks, the node's degree is highly unequally distributed. Panel \figref[B]{grid_degree_wo_distribution}  highlights this by depicting the share of nodes ordered by their degree on the x-axis and the share of the accumulated total degree on the y-axis. Put differently, the bottom x-percentage of the degree distribution account for the corresponding y-percentage of total degree and if all nodes would have the same degree, the line would equal the bisecting, dotted line. While we observe inequality in degree in all layers, these are most substantial in the colleague and classmate layer, in which the top 20\% of the distribution accounts for roughly half of all connections. Panel \figref[C]{grid_degree_wo_distribution}  underlines that the distributions are heavy-tailed when stacking up the layers to a single network: a large share of the overall distribution is accounted for by relatively few, highly central nodes. With this concentration of degree our network resembles results that are also found in other realized \cite{stopczynski_measuring_2014} or administrative \cite{bokanyi_anatomy_2023} social networks.

\begin{figure}
    \centering
\includegraphics[width=1\textwidth]{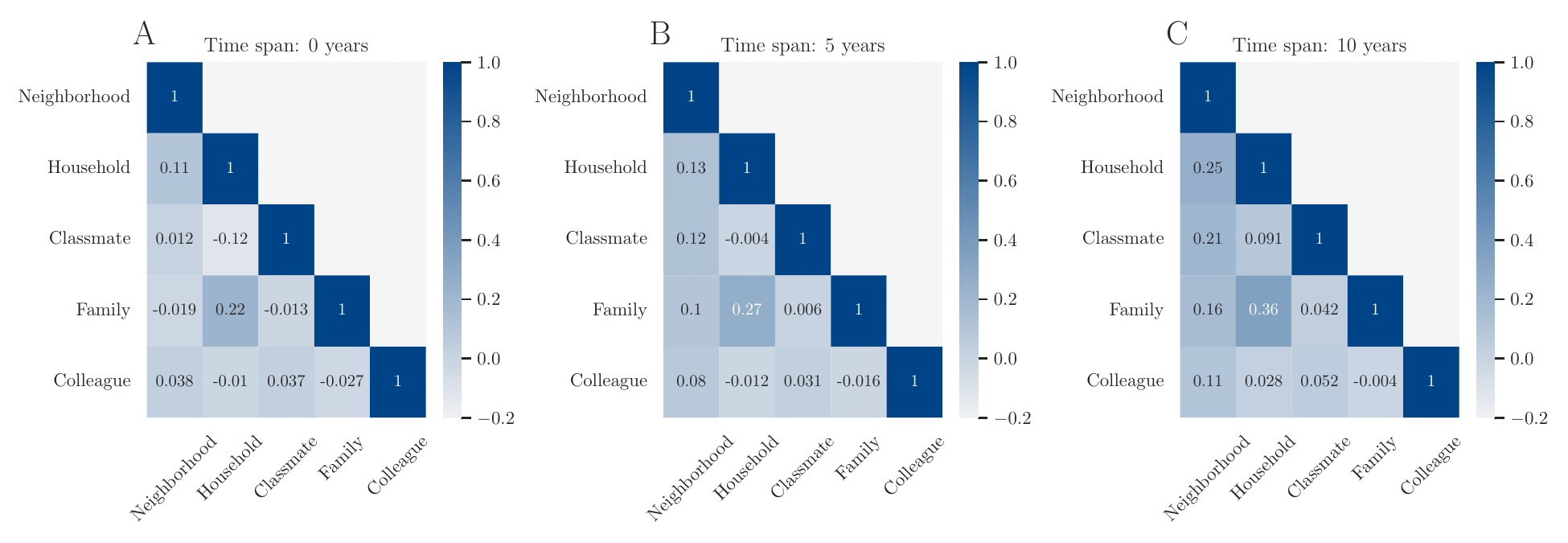}
\caption{ \textbf{Inter-layer Degree Correlation.} 
Pearson correlation between node's degree for varying layers and time-span. The time-span applies to the layers on both the x- and y-axis. Panel A shows the correlation with no time-span, Panel B with time-span of five years and Panel C with time-span ten years.}
    \label{fig:degree_correlation}
\end{figure}

Next, we analyze whether the fat-tailed, stacked degree distribution is driven by single layers or whether individuals having a large number of connections in one layer directly implies that those individuals also have comparably more connections in other layers.  Figure \ref{fig:degree_correlation} depicts the interlayer degree correlation for different time-spans. We observe a positive correlation of degree in most layers with a substantial increase in the magnitude of correlation with rising time-span across layers. The correlation is especially pronounced between the family, household, and neighborhood layers. The increase in correlation coefficients over time-span indicates some preferential attachment, i.e., well-connected nodes on some layer are more likely to have more edges on other layers and this effect accumulates over time \cite{barabasi_emergence_1999}. 

\begin{figure}
    \centering
\includegraphics[width=1\textwidth]{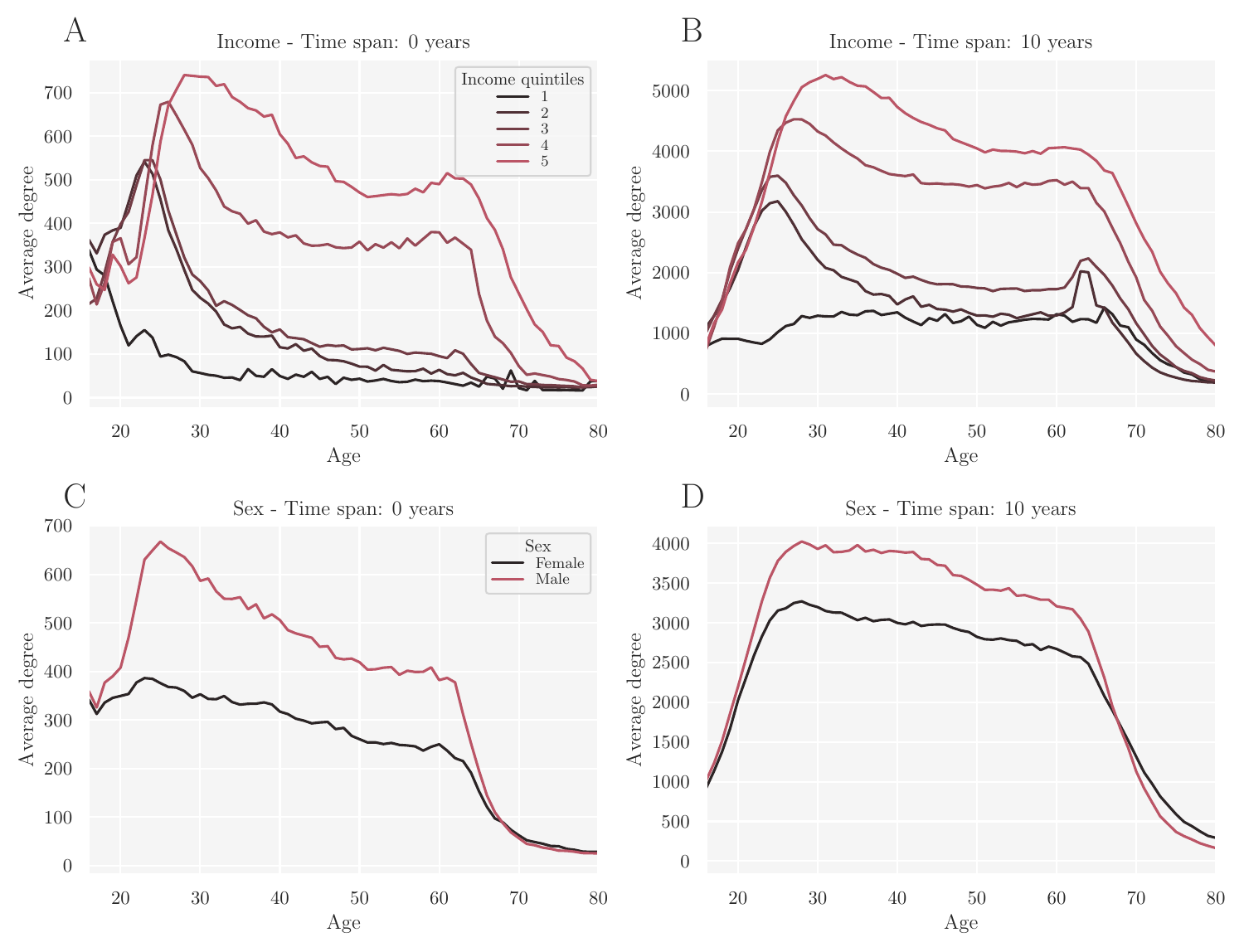}
    \caption{ \textbf{Relation Of Individual Degree And Income Or Sex Over The Working Life Cycle.} 
    Average degree distribution conditional on age and income (Panel A and B) or age and sex (Panel C and D).
     The plot is based on 2021 data. Panels A and C contain a single year time-span andPanel B and D a ten year time-span.
   }
    \label{fig:degree_age_income}
\end{figure}

We also investigate how the degree as an indicator for social opportunities reflects life outcomes\cite{bokanyi_anatomy_2023}. Figure \ref{fig:degree_age_income} shows how the average degree sorts closely by the income for a given age. The degree-differences by income are especially pronounced for persons in the middle of their life between ages 25 and 60 and when taking past connections into account (Figure\figref[B]{degree_age_income}). It has also been suggested in the literature that women have a lower average degree than men, leading to labor market differences \cite{lindenlaub_network_2021}. Figures\figref[C]{degree_age_income} and\figref[D]{degree_age_income} show that in our network degree indeed differs by sex when conditioning on age for most of the working life cycle and reverses slightly for persons in the pension age. While neither of these findings imply any direction of causality, it underlines how closely the network's centrality and socioeconomic trajectories interrelate. Additionally, it demonstrates the potential of integrating further individual-level data from registries into the network.

\subsection{Higher-order clustering}
Because relations form via shared foci, most network layers are highly clustered by design. Therefore investigation of clustering only begins to reveal socially relevant information when considering connections spanning multiple years or considering multiple layers at the same time. Beyond the tendency of nodes to form triangles, the importance of higher-order structures in complex (social) networks has been emphasized in recent literature \cite{benson_higher-order_2016, bick_what_2023,yin_higher-order_2018}. This also includes the investigation of higher-order clustering coefficients \cite{yin_higher-order_2018}. The $\ell$-order clustering coefficient $C_{\ell}$ measures the probability of an $\ell$-sized cliques and a randomly drawn adjacent node to form an $\ell +1 $ clique. For $\ell=2$ the higher-order clustering coefficient simply equals the canonical clustering coefficient. To give an example from our network, consider the classmate layer and the third-order clustering coefficient. The local clustering coefficient for a person would give the probability that of three of its classmates a randomly drawn further neighboring node would also share that class. It is thus a measure of the relevance of cliques of different sizes. A visualization of the higher-order clustering coefficient can be found in Appendix \ref{sec:appendix-vis-hocl}. 

Figure \ref{fig:hocl} depicts the distribution for higher-order clustering for cliques of two and three when combining all layers for a single year using the bipartite view. We observe a large share of nodes with almost complete higher-order clustering for both $\ell$ (7.1\% (10.3\%) of nodes have a clustering coefficient larger than 0.95 for cliques of size two (three)). For $\ell=3$, the distribution of clustering coefficients shifts right and the average local clustering coefficient increases, going against the observations in multiple artificial and real-world networks \cite{yin_higher-order_2018}. This result indicates the high relevance of larger highly-clustered groups in our network. It thus underlines how not only the individual layers but also the network of combined layers can be understood as one of foci group affiliation, adding an additional rationale for the bipartite perspective. In Appendix \ref{sec:appendix-hocl}, we present how we utilized the bipartite structure to compute the higher-order clustering efficiently at scale.

\begin{figure}
    \centering
\includegraphics[width=1\textwidth]{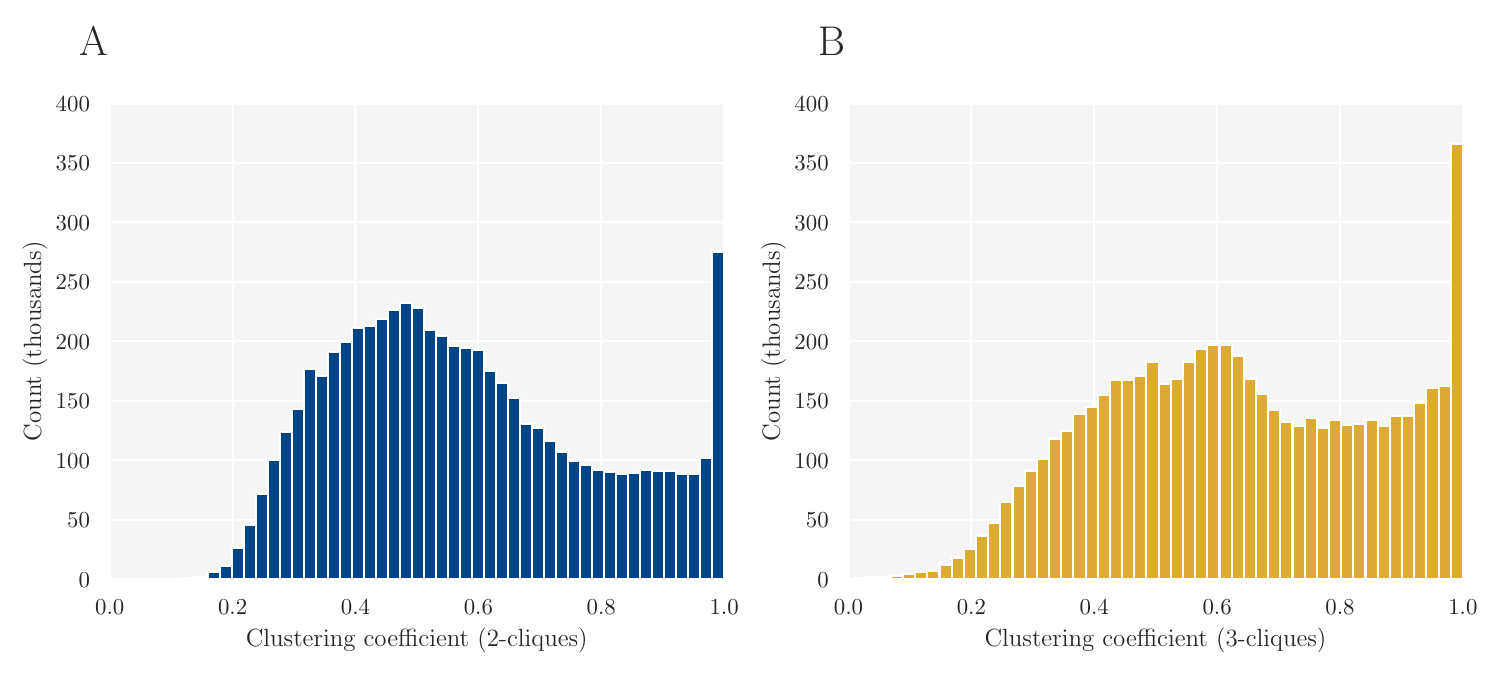}
    \caption{
    \textbf{Higher-Order Network Clustering.} 
Density of higher-order local clustering coefficients. Panel A shows for cliques of size 2 and  B for size 3.} 
    \label{fig:hocl}
\end{figure}

\subsection{Shortest paths distributions} \label{sec:shortest_paths}
The average distance between a node to all other nodes is an important measure for social distance, information transitivity, or disease spreading \cite{stopczynski_measuring_2014, guilbeault_topological_2021,travers_experimental_1969,watts_new_2004}. In large-scale networks, computing distances between nodes becomes increasingly computationally expensive as the complexity for commonly used search algorithms (e.g., the Dijkstra or Johnson algorithm) grows both in the number of nodes and edges \cite{fredman_fibonacci_1984,johnson_efficient_1977, dijkstra_note_1959}. We therefore sample shortest-path length distributions by computing the distances between 1000 nodes chosen uniform at random from the node set to 1000 other nodes chosen in the same fashion, shown in Figure \figref[A]{shortest_path_grid} for different layers and time-spans. Note that we exclude pairs that are located in separate, unconnected components of the network.

Comparing the individual layers, we observe different magnitudes and effects of time-span: The family layers' average distance hardly changes over time and connects a large fraction of the population. Within the neighborhood, classmate and colleague layers, the addition of past connections decreases the average distance between individuals for larger time-spans due to a higher number of bridges between the locally clustered connected components. Households are relatively stable, small-scale and completely clustered foci. Therefore paths exist only in large samples and all nodes in our sample are unconnected except for a single pair until a ten year time-span.

\begin{figure}
    \centering
\includegraphics[width=1\textwidth]{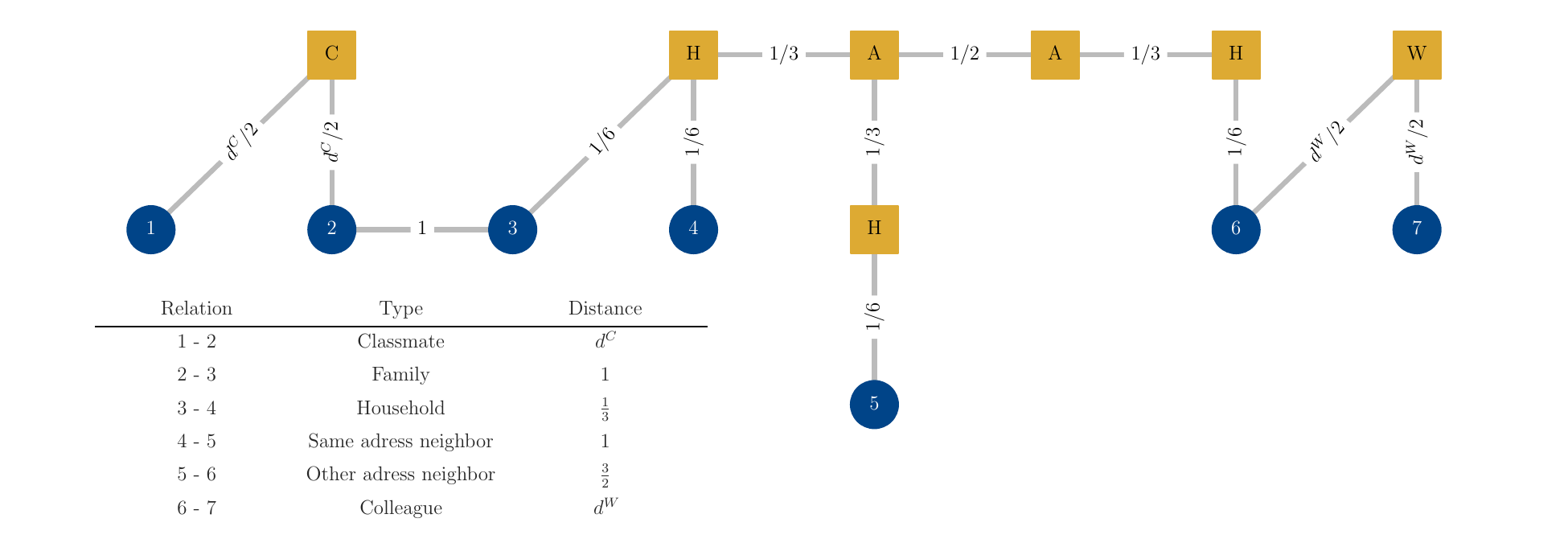}
\caption{
\textbf{Bipartite Representation and Edge Weighting Scheme.}
Visualization of the suggested weighting scheme for meaningful distances. Individual nodes are depicted as blue circles and container nodes as yellow squares with the number or sign on the nodes denoting the individuals' pseudo id for visualization purposes or the abbreviated container type (``C'' for classroom, ``H'' for household, ``A'' for address, ``W'' for workplace). The number or sign on the edge indicates the suggested weight. The distances between colleagues and classmates are adjusted to the degree of the container node, i.e., the size of the workplace or class (for details see Appendix \ref{sec:appendix-robustness-shortest-paths}). Note that all of these weights can be customized to the research question at hand using the \emph{regnet} Python package. }
    \label{fig:weight_visualisation}
\end{figure}

For many research questions in social science and beyond, combining multiple layers is important. This is because instead of isolated social foci, researchers are interested in the joint social interaction space, e.g.\ when attempting to trace where an individual was infected with a virus. The different scales and temporal patterns of average path lengths and the variations in degree suggest that the likelihood that an edge in a given layer represents an actual social interaction is not equal for each layer. Treating node-distances the same for every layer would mean that an individual would have the same distance to a colleague in a workplace with a thousand employees as to their child. Therefore, edge-weights between individuals should be adjusted to reflect the distinct nature of the layers. In the following, we utilize the bipartite view to suggest a framework of edge-weights with universally intuitive properties, as visualized in Figure \ref{fig:weight_visualisation}.
For the network of combined layers to reflect meaningful distances, we set the distance between individuals as a reference with unit length one. We gauge the other layer's edge-weights against that. We assume household connections to be the strongest as interactions in households are highly likely to occur frequently. We therefore set the weight of persons to the household container node to $1/6$, so that two persons in the same household have distance $1/3$. In the bipartite view of the network, neighborhoods are connected to each other via address containers. We set the weight between the address containers to $1/3$ so that two individuals living at the exact same address but not forming a household  (e.g., roommates in a dorm) have distance $1$, thus the same distance as family members. We assume neighbors living at a different address to have a lower interaction likelihood and therefore set the weight between addresses to $1/2$ so that neighbors have distance $3/2$. We furthermore assume that the likelihood of realized interactions at the workplace and in school programs decreases with the number of colleagues or classmates and therefore adjust the weights for work connections by the size of the company. A more detailed description of the weighting for the colleague/classmate layer can be found in Appendix \ref{sec:appendix-robustness-shortest-paths}. 

\begin{figure}
    \centering
    \includegraphics[width=0.82\textwidth]{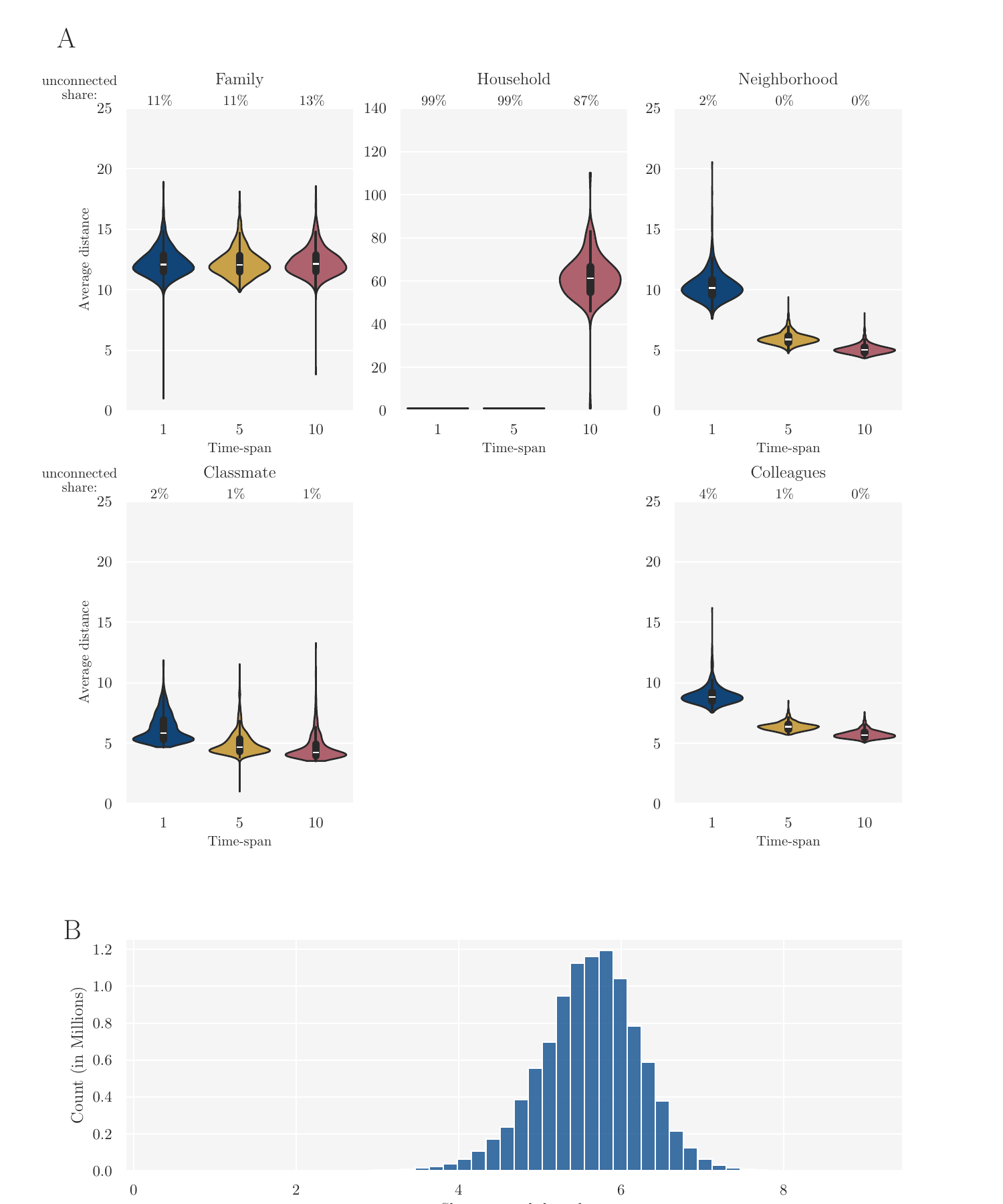}
    \caption{
    \textbf{Shortest Path Lengths.} 
    Panel A illustrates the distribution of the average shortest path for a sample of 1000 random nodes to another 1000 random nodes for different layers, and time-spans (x-axis and color). The number above the plots indicates the share of unconnected nodes in the sample. Panel B shows the distribution of 1 million random distances in the bipartite network with all layers combined. The edge-weights are adjusted to reflect meaningful relationships between the layers.}
    \label{fig:shortest_path_grid}
\end{figure}

We compute the average distance between a random sample of a thousand individuals to another thousand individuals in this network. The resulting distribution (see Panel B of figure\ref{fig:shortest_path_grid}) is sharply-peaked with an average distance of 5.6, which aligns with the "six degrees of separation" found in many social networks, e.g., online platforms \cite{ugander_anatomy_2011} or communication networks \cite{stopczynski_measuring_2014, leskovec_planetary-scale_2008}. Note that the weights we used are just one suggestion that leads to intuitive lengths. Weights can be easily tailored to domain knowledge for a given research purpose. An analysis of the robustness of the suggested weights can be found in Appendix~\ref{sec:appendix-robustness-shortest-paths}. 

\subsection{Software and Python package}\label{sec:software}

 For the construction of the bipartite view and to compute relevant network metrics, researchers get access to the \emph{regnet} Python software \cite{maier_regnet_2024-1}. The package is based on the Python interface for the C library \emph{igraph}, which is also used to create graphs and compute network metrics for the individual-centered view. More details on the \emph{regnet} software can be found in Appendix~\ref{sec:appendix-regnet} and the repository \cite{maier_regnet_2024-1}.

\section{Conclusion}
In this paper, we have presented a new, nation-scale network dataset derived from Danish registry data containing relationships for multiple central levels of social interactions. Unlike any other published administrative network dataset, this network spans more than a dozen years and is accessible in a dual view -- an individual-centered view that only contains individuals and their relations and a bipartite view including container nodes at which individuals cluster. We have shown that the bipartite view is conceptually plausible because of the high clustering of nodes around shared foci and results in efficiency gains due to the drastically reduced number of edges. We furthermore introduce an edge-weighting scheme that makes it possible to connect multiple layers while maintaining plausible distances between individuals. This weighting of edges within the bipartite view is implemented in a comprehensive Python package that is published alongside the dataset. Looking at its central characteristics, the network incorporates further insightful yet intuitive social properties. These include the heterogenous, heavy-tailed distribution of degree, a time-span increasing inter-layer degree correlation, small but relevant overlaps between different layers, as well as the striking relationship between an individuals' number of connections and their relative income position within the same age group.

By making the network accessible to the research community, we are confident that this dataset can play a central role in a wide range of fields. For many research questions, such as the study of social mobility, examining the relationship between network structure and individual-level outcomes is at the core of the inquiry\cite{chetty_changing_2024,chetty_social_2022,chetty_social_2022-1}. 
 Our research, together with existing research in the social sciences, points to potential path-breaking possibilities of combining individual-level administrative data with our network. First, research could analyze nation-scale network data over longer time horizons that would enable tracking of social cohesion, e.g., by measuring the connectedness of the social network across time and sorting by individuals' sociodemographic attributes. Second, research could combine the rich administrative data on individuals' connections and characteristics with survey data to estimate micro-level relational data \cite{breza2020using, breza2023consistently, alt_diffusing_2022}. Third, further studies could utilize the bipartite view and additionally attach administrative attributes to the container nodes to investigate the relation of institutional and individual influences on link formation. Finally, new research could use the social measure to understand peer effects and social influence \cite{kassarnig_class_2017} using accurate individual measures that are often not available to social media companies. The administrative data also allows researchers to establish causal social effects, e.g., externalities in the domains of consumption \cite{de_giorgi_consumption_2020,tebbe_peer_2022}, family and health behavior \cite{dahl_peer_2014,fadlon_family_2019}, political concerns \cite{alt_diffusing_2022}.

\clearpage

\bibliography{references}

\clearpage
\section*{Appendix}
\appendix
\section{\emph{regnet} python package}
\label{sec:appendix-regnet}
The \emph{regnet} package is divided into two central data structures. The first is a simple single year, unweighted, multi-layer structure suitable for determining connections, neighborhoods, and degree. The second structure is multi-layer, can span multiple years, and allows edge-weights to be tailored to the research question at hand. A proposal for a weighting scheme that reasonably resembles interaction probabilities is presented in Subection \ref{sec:shortest_paths}. This structure is designed to efficiently find shortest paths between nodes, and presents the layer types through which they are created. A detailed instruction guide on how to use the different structures and methods is provided in the \textit{Readme} section of the \emph{regnet} repository \cite{maier_regnet_2024-1}.

\section{Visualization of the local higher-order clustering coefficient}
\label{sec:appendix-vis-hocl}

\begin{figure}[H]
    \centering
\includegraphics[width=\textwidth]{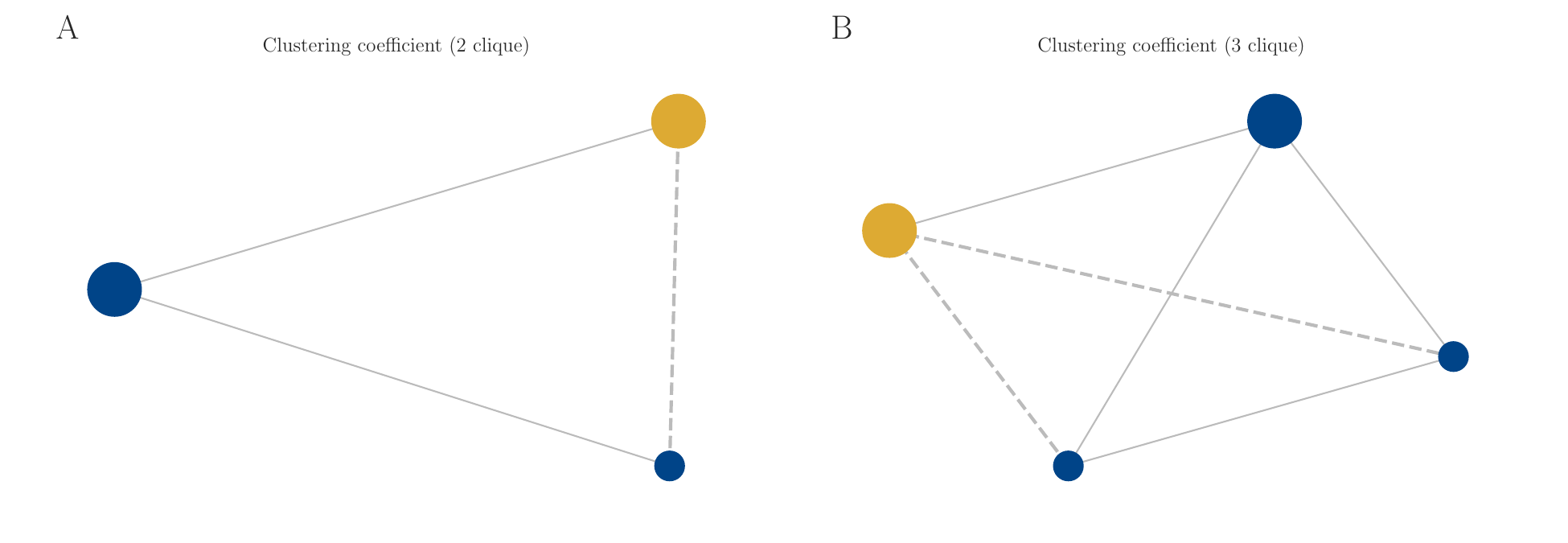}
\caption{
    \textbf{Visualization of the local higher-order clustering coefficient \cite{yin_higher-order_2018}.}
    The blue nodes denote nodes belonging to a clique. The local clustering coefficient of the large blue node can be computed as the share of adjacent edges (as the one to the yellow node) that also have edges to all the other nodes in the clique (i.e., the dotted edges) relative to all adjacent edges.
    }
    \label{fig:visualisation_hocl}
\end{figure}

The intuition between the higher order clustering coefficient is illustrated in  Figure \ref{fig:visualisation_hocl}.
The blue dots denote nodes belonging to a clique of size two (\figref[A]{visualisation_hocl}) or three (\figref[B]{visualisation_hocl}). The higher-order local clustering coefficient of the large blue node gives the probability of an adjacent node -- as the yellow one -- to be connected to all the other nodes in the clique, i.e., the probability that the dotted edges exist \cite{yin_higher-order_2018}.  

\section{Computation of the local higher-order clustering coefficient} 
\label{sec:appendix-hocl}
Higher-order clustering can be efficiently measured in the bipartite representation of a network with container nodes.
The $\ell$-th order local clustering coefficient for a node $u$ can be computed via
\begin{gather}
    C_{\ell}(u)=\frac
            {\ell \lvert \mathcal K_{\ell+1} (u)  \rvert}
            {(d_u-\ell-1)\lvert \mathcal 
 K_{\ell}(u)\rvert}
\end{gather}
where $d_u$ denotes the degree of $u$ and $\mathcal  K_{\ell}$ the set of $\ell$-cliques containing $u$ \cite{yin_higher-order_2018}. 
In the bipartite network, we compute the person-degree of $u$ by taking the length of the set of all person-neighbors in all layers. To compute the total number of the $\ell$-sized cliques containing $u$, it is key to realize that all individuals connected to a container node are completely clustered, i.e., they are all connected to each other. The number of cliques created via a container node can, therefore, be computed using simple combinatorics.
\begin{gather}
        \left \lvert
            \mathcal K_{\ell}^{\mathcal C_i}(u)
        \right \rvert
        =
         \genfrac(){0pt}{}{\lvert {\mathcal C_i} \rvert -1}{\ell-1}
\end{gather}
where $\mathcal C_i$ is the set of person-neighbors of a  $u$-neighboring container node. There is a correction by $(-1)$ because we want to obtain the number of cliques containing $u$, not all cliques in $\mathcal C_i$. As cliques can overlap between different layers and containers (e.g., a person can be both members of the same household and family or be a colleague at two different firms), we have to correct for the overlaps using the principle of inclusion-exclusion when computing the total number of $\ell$-sized cliques. 

\begin{gather}
        \left \lvert \mathcal K_{\ell}(u) \right \rvert
        =
        \sum_{\mathcal S\subseteq \{1,...,\vert \mathcal C\rvert \}}
        (-1)^{|\mathcal S|+1} \genfrac(){0pt}{}{\lvert\cap_{s \in \mathcal S}{\mathcal C_s} \rvert -1} {\ell-1}
\end{gather}
where $\mathcal C$ is the set containig all $\mathcal C_i$'s. 

\section{Weighting scheme}
\label{sec:appendix-robustness-shortest-paths}
\subsection{Computation of weights for colleague and classmate layer}
To compute the edge-weights for the colleague and classmate layer, we assume a $k$-regular-random network structure, i.e., every colleague/classmate is connected to exactly $k$ other random colleagues/classmates. For this, we need to define how many individuals $k$ an average individual is connected to through their class or workplace container. A suitable $k$ in our case is $k=20$, roughly corresponding to the long-term average classroom size in Denmark \cite{dst_primary_2023}. 
While assuming random, independent formation of edges is comparatively unrealistic, the average shortest path length of a $k$-regular random networks scales similarly to those of small-world networks of more complex, realistic structures, e.g., Watts-Strogatz or modular hierarchical networks \cite{watts_collective_1998,watts_identity_2002}. Importantly, the $k$-regular-random network provides a useful equation for analysis, intuitively relating increased degree of connectivity $k$ to shorter path lengths. Consider an employer/education program size $n$. For every employer of size $n$, the average $k$-random-regular distance is defined as: \begin{align*}
    d=\mathrm{max}\left(1,
           \frac{\log(n)-\gamma}{\log(k-1)}+ \frac{1}{2} + \frac{\log(1-2/k)}{\log(k-1)}\right),
\end{align*} where $\gamma$ is the Euler-Mascheroni constant. We set the weight as the maximum of the average distance in the $k$-regular-random network and $1$ to assure that in small companies the distance between colleagues is of unit length. In a company of 500 employees the estimated distance between two individuals is then $2.13$, for the largest workplace present in the dataset it is $3.2$.

\subsection{Robustness checks}
\begin{figure}
    \centering
\includegraphics[width=\textwidth]{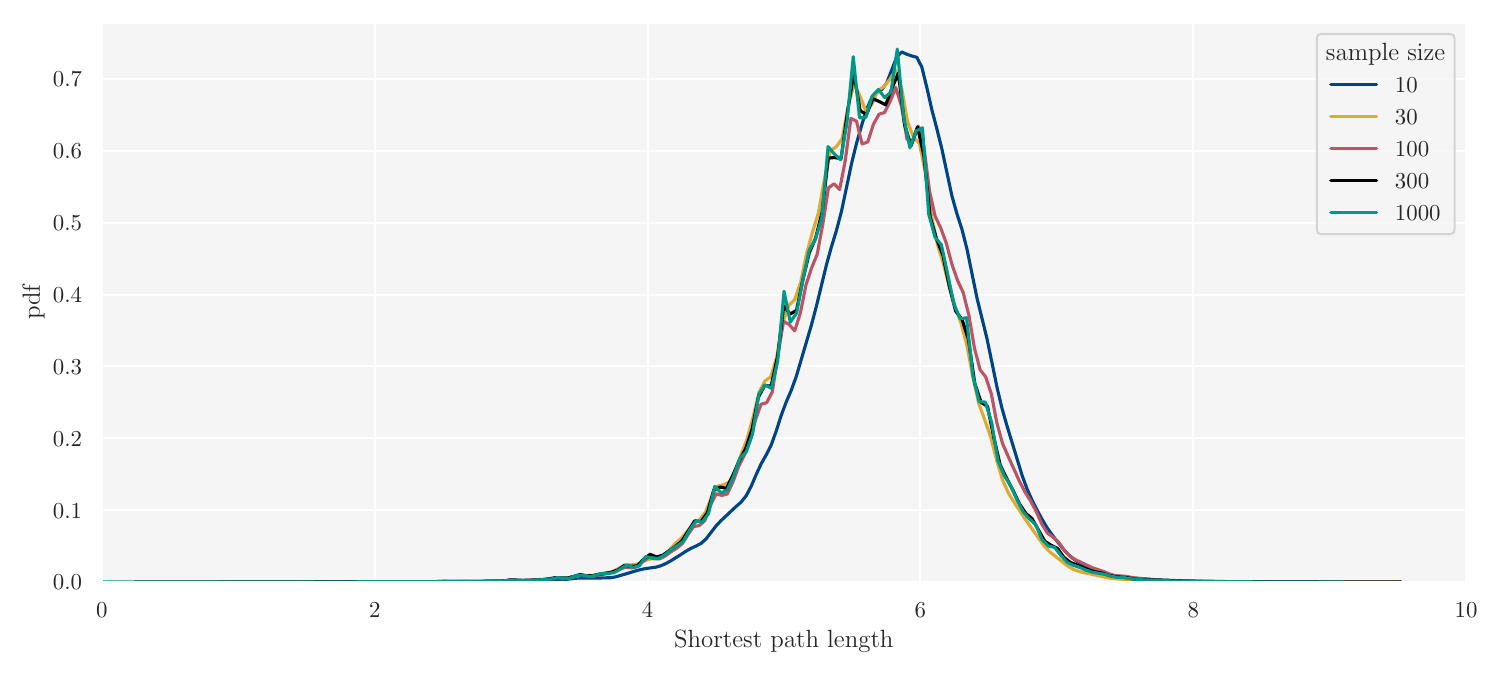}
\caption{
    \textbf{Shortest Path Distribution with Varying Sample Sizes.}
Density of the distribution of shortest paths of a varying number of random nodes to 10000 other random nodes in a single year network.}
    \label{fig:robustness_sample_size}
\end{figure}

\begin{figure}
    \centering
\includegraphics[width=\textwidth]{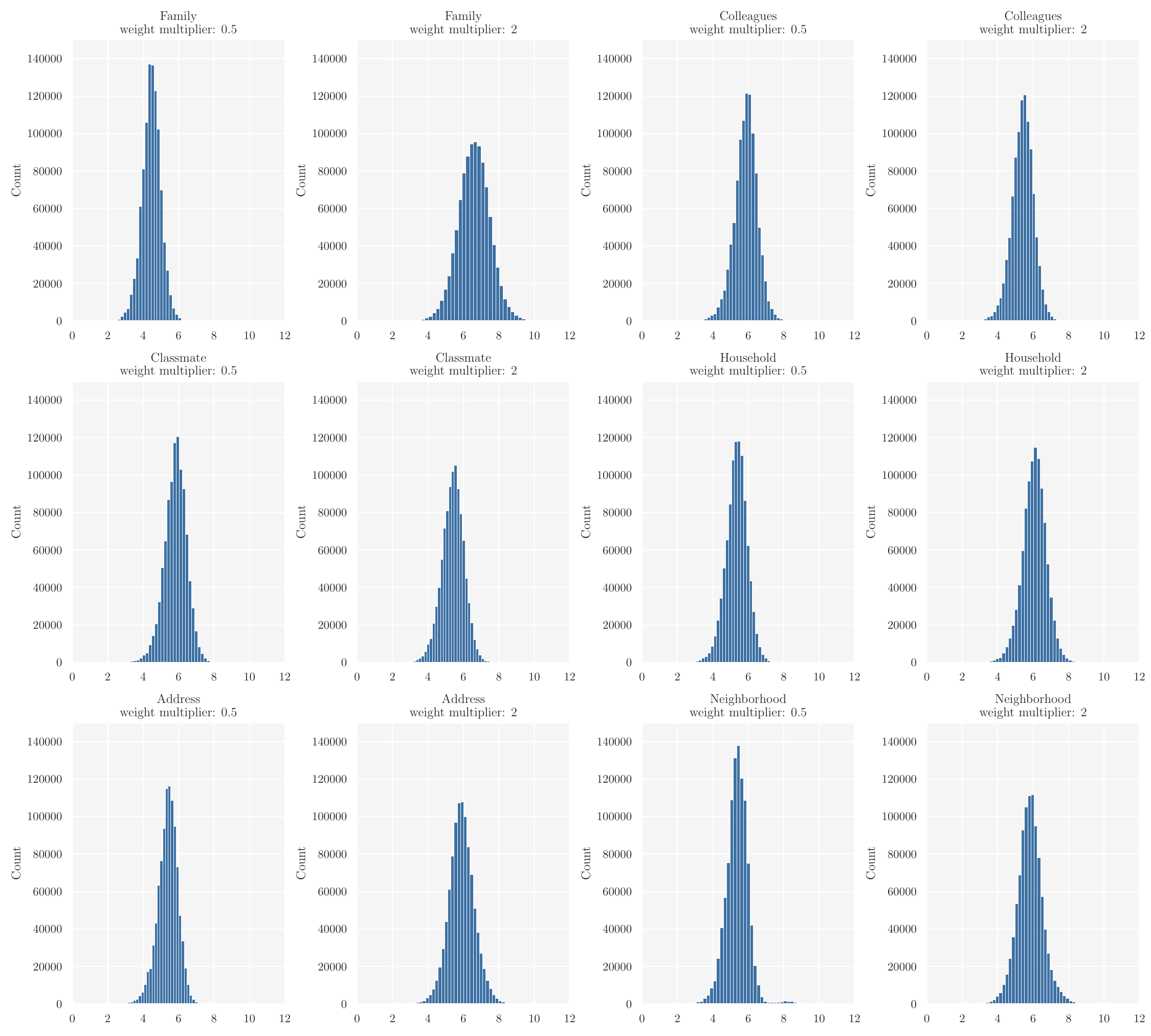}
\caption{
    \textbf{Shortest Path Distribution with Varying Edge Weights.}
Distribution of shortest paths from $100$ random nodes to $1,000$ other random nodes in a single year network, varying the weights on a single layer and leaving the other edge-weights as described in the Subsection \ref{sec:shortest_paths} and Subection \ref{sec:appendix-robustness-shortest-paths}. The title of each plot indicates the layer with changed edge-weights and the factor with which the weights described in the above mentioned sections are multiplied (left plot: half the weight, right plot double the weight).}

    \label{fig:robustness_weight_changes}
\end{figure}

\begin{figure}
    \centering
\includegraphics[width=1.0\textwidth]{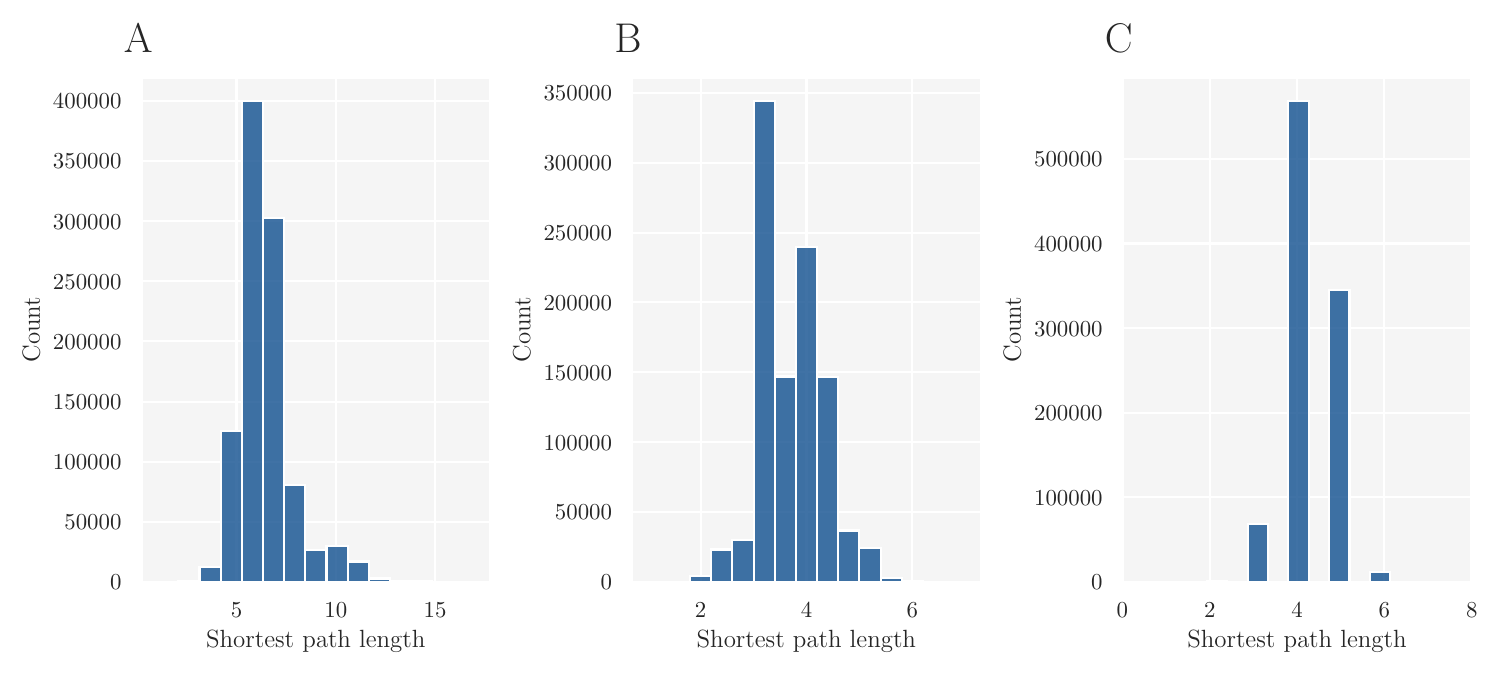}
\caption{
    \textbf{Shortest Path Distribution With Naive Weights.}
Distribution of shortest paths from 100 random nodes to $1,000$ other random nodes, when not assigning any weights in the bipartite view(Panel A) or assigning uniform weights to the colleagues and classmate layers of $1/2$ in the bipartite view (Panel B). Not assigning any weights in the bipartite view implies that neighbors have distance 5 (see Figure \ref{fig:weight_visualisation}, while setting uniform distance means that the distances between colleagues and classmates are 1 independent of the size of the workplace or class. Panel C illustrates not assigning any weights in the individual-centered view (with sampling for larger workplaces and classes).  }

    \label{fig:robustness_unweighted_uniform}
\end{figure}

Figure \ref{fig:robustness_sample_size} illustrates the reasonability of our sampling approach for shortest paths. The plot presents the distribution as a kernel density function of a varying number of source nodes to $10,000$ random target nodes. We observe quick convergence of the distribution with increasing size of sampled source (target) nodes (cf. Figure \ref{fig:shortest_path_grid}).
Figure \ref{fig:robustness_weight_changes} depicts changes in the distribution of shortest paths when the edge-weights of individual layers are either halved or doubled and all other weights are left constant compared to the scheme suggested in Subection \ref{sec:shortest_paths}. The adjustments of the weights lead to minor shifts in both the overall distribution and increased dispersion, especially for the family layer. Overall the distribution remains in a reasonable range compared to other social networks \cite{stopczynski_measuring_2014, leskovec_planetary-scale_2008, ugander_anatomy_2011}. The robustness check demonstrates the necessity to adjust the weights to the research question at hand.
Figure \ref{fig:robustness_unweighted_uniform} demonstrates the distribution of distances in the bipartite network when there are either no weights (\figref[A]{robustness_unweighted_uniform}) or distances between classmates and colleagues are not adjusted for their size (\figref[B]{robustness_unweighted_uniform}). With all edge-weights equal to one the distances become more concentrated. Note that here, almost none of the shortest paths lead through the neighborhood layer as not setting any weights implies a distance of 5 between neighbors. Adjusting the classmate and colleagues weights to uniform length leads to significantly smaller average shortest paths. Similarly, the distribution of distances when not adjusting any weights utilizing the individual-centered view becomes more concentrated around 4 and smaller on average (mean distance in the random sample is 4.3) (\figref[C]{robustness_unweighted_uniform}). Compared to other social networks these are implausibly small, underlining the need for a meaningful adjustment of weights.

\clearpage
\section{Detailed layer definitions}
\label{sec:appendix-layers}

\subsection{Family layer}
The family relations from which this layer is created are recovered from the Danish population registry \cite{petersson_danish_2011} which includes all individuals alive and living in Denmark from 1968 onward. This registry is augmented with archival data on marriages and births from the Danish national archives to reconstruct as many relations between parents and children alive in 1968 for individuals born in or after 1953. 

All family relations are extrapolated from the parent-child relations in the registries. If a new family relation is created, e.g.\ via the birth of a child, this relation is included in the edge list for a specific year independently of the day of the year the relation is created. Note that to recover as many relations as possible deceased individuals were included in the data, as well.

Concerning coverage, all individuals in the data born in 2019 and 2020 have known parents and the percentage of individuals with known parents is over 73\% for all individuals born in 1952 or later. For individuals born before 1952 the pro-portion of individuals with known parents varies from 10\% for those born in 1940 to 27\% for those born in 1951.

\subsection{Household layer}
Statistics Denmark computes a household identification number using information on address, age, and marital status from the population registry. The household information included in the household layer concerns households as registered on the first of January in a specific year. We do not allow for individuals to belong to more than one household in the same calendar year. The household identification number used in the current version of the network is available from 1986 and is updated every three months. Individuals in a household live at the same address and a household consists of an individual or a couple with our without children. 

Children have the same household identification number as their parents if
\begin{itemize}
\item they live at the same address as at least one of the parents,
\item are under 25 years of age,
\item have never been married or in a registered partnership,
\item don't have children of their own that are registered in the central person registry (CPR),
\item are not part of a cohabiting couple.
\end{itemize}

A couple is defined as two individuals that live together and are one of the following types:
\begin{itemize}
\item married couple
\item registered partnership (from October 1st, 1989)
\item living together and have at least one child that is registered in the CPR
\item cohabiting defined as two individuals of opposite sex with an age-difference of under 15 years that do not have any children registered in the CPR and are not close family (brother-sister / parent-child) according to the CPR. In addition, the household has to contain only adults.
\end{itemize}

Children not living at home are under 18 years of age and each make up their own family. Technically they are counted as single. To be considered a child not living at home the following conditions need to be fulfilled:
\begin{itemize}
\item don't live at the same address as either of their parents
\item be under 18 years of age
\item never have been married or in a registered partnership
\item do not have children registered in the CPR
\item are not part of a cohabiting couple
\end{itemize}

Rules for changes in family identification number over time:
\begin{itemize}
\item When individuals in a couple move apart they each get a new id.
\item When two individuals move together and create a couple they both get the same new id.
\item When children move out of the home they get their own new id but the id for the rest of the family does not change.
\item When children that previously moved out move back with their parents again before they turn 25 they get the same identification number they previously had.
\item When one individual in a couple dies, the surviving part and their children living at home still get a new id. The deceased individuals children that are not also children of the surviving individual each get their own new id.
\item When the couple type (see above) changes or the couple separates but still live together, the family id does not change.
\end{itemize}

Identification numbers that have been decommissioned are not re-used. Note that in some cases multiple identification numbers are registered for the same address. This is for example in case of three adults living at the same address where none are married or cohabiting. But also institutional households such as college dorms or care homes are examples of this.

Disambiguation: Note that the household definition used in the current version of the network corresponds to what in layman’s terms may be regarded a family. 
The definition of a household used at Statistics Denmark may thus lead to cases in which one would expect individuals to be members of the same household but are not labelled as such according to our definition: e.g. children over 25 living at home do not form a household together with their parents and roommates do not form a household either. In the first case an edge between parent and child will however be present in the family layer, and in both cases an edge is likely in the neighborhood layer. 

\subsection{Neighborhood layer}
In the neighborhood layer we include edges between an individual and the members of the geographically closest $10$ households on the first of January of a specific year (as defined in the household layer) within a distance of $50$ meters. This is the same approach as \cite{van_der_laan_whole_2023} take. Households for which the exact location is not known are disregarded at present. When multiple households exist with the same distance, households are selected uniformly at random. Note that when households are sampled this does not automatically lead to symmetry in the sense that if individual $i$ is sampled as neighbor to individual $j$, individual $j$ is not necessarily sampled as neighbor to individual $i$. 

Note that the above definition implies that in sparsely populated areas, there may be under 10 households within a 50 meter radius. It may be worth considering and it is possible using the bipartite data to define other distances in the network to reflect other definitions of neighborhood. The address and exact geographical location of almost every household is known and included in the bipartite table for the neighborhood layer.

Disambiguation: Note that since a household in the network is what in layman’s terms may be seen as a family, the neighborhood may for some individuals easily capture mainly other families in the same household, since their distance is 0 meters. E.g. for a person living in a collective, with 11 individuals, with no children and no married couples will not have any neighbors that are not part of the same address, as the individual's $10$ closest households will all have distance zero. So in the colloquial meaning of the word ``neighbor'', such a person would have zero neighbors. Note also, that apart from such extreme edge cases, even $4$-person collectives will then have at most $7$ neighbors in the colloquial sense, even though the person will have $10$ neighbors in the neighborhood-layer.

\subsection{Colleague layer}
In this layer, we use the information on work location from the labor market registry \cite{petersson_danish_2011}. This registry contains a yearly status indicating the place(s) of work on the 30th of November of all individuals present in the Danish population on the 1st of January of the same year. For small workplaces ($\leq100$ employees) we include edges between all employees in the yearly edge-list for the individual-centered view. For large workplaces ($>100$ employees) we take a random sample of size $100$ to create edges between colleagues. We choose such a limit to make sure that the edge lists do not become too large (the amount of edges grows at a rate of $\propto n^2$, and because it is unrealistic that an individual actually meets all colleagues at a large workplace. The network file can however be adapted to adjust the workplace size boundary that determines whether we sample colleagues or not. Note that when individual $i$ is sampled as a colleague for individual $j$, we make sure to include an edge in the other direction as well, such that individual $j$ is included as a colleague for individual $i$. This condition leads to some individuals from sampled workplaces having less than 100 colleagues (especially for those with < 200 employees). A final thing to consider regarding the sampling is the fact that if one is sampled as a colleague in one year this does not automatically mean one is also sampled for other years (even if both individuals still work at the same work-place). 
For the bipartite-view we do not sample as here the number of edges only grows linearly with the number of colleagues compared to the quadratic growth in the individual-centered view. 
Finally note that individuals that have a so-called `fictional workplace', meaning individuals that work remotely or at multiple locations in the same municipality, are count-ed as having no colleagues. Independent business owners and spouses involved in their businesses are not included either.

\subsection{Classmate layer}
In this layer we use information on enrolled students in primary, secondary, and tertiary education from the Danish education registers \cite{jensen_danish_2011} to create edges between students enrolled at the same school for the same program in the same grade or study (start) year. The data cannot be used to distinguish between multiple classes within the same grade and program. The data contain information from the 1st grade of primary school (from the year an individual turns six) onward. Note that an edge is created independent of how much overlap (in days) there was during the enrolled period. Also individuals that were enrolled in the same grade and program for just one day are considered classmates. 
For the bipartite-view, programs are included as container nodes to which individuals are connected.

\section*{Author contributions statement}
J.C.: designed the network and supervised its creation,
B.K.: responsible for the description and analysis of the network,
B.F.M.: Developed the bipartite view of the networks including the edge-weighting scheme and wrote the \emph{regnet} Python package,
S.N.E., F.C.: Assisted in the creation of the network,
S.L.J., A.B.N., L.H.M., D.D.L.: conceived the original idea,
B.K., J.C., J.E., L.H.M., A.B.N.: wrote the paper with input from all authors
A.B.N.: supervised the description and analysis

\section{Data availability statement}
The data and the used software are available for qualified researchers via Statistics Denmark (see  Subsections~\ref{sec:access} and ~\ref{sec:software}).

\section*{Additional information}
This paper replaces a previous version presenting the dataset \cite{bjerre-nielsen_dataset_2024}.

The authors declare no competing interests. DDL is deputy chairman of the Board at Statistics Denmark.

\end{document}